\documentclass{aa}  

\usepackage{graphicx}
\usepackage{txfonts}
\usepackage{mathrsfs}
\usepackage[colorlinks,allcolors=blue]{hyperref}
\usepackage{breakurl}
\usepackage{subfigure}

\def \src {\mbox{A0538$-$66}}
\defcitealias{Alcock01}{A01} 	
\defcitealias{McGowan03}{MC03} 
\defcitealias{Charles83}{C83} 
\begin{document}

   \title{Optical and near-infrared photometric monitoring of the transient X-ray binary A0538$-$66 with REM}


   \author{L. Ducci
          \inst{1,2}
          \and
          S. Covino\inst{3}
          \and
          V. Doroshenko\inst{1}
          \and
          S. Mereghetti\inst{4}
          \and
          A. Santangelo\inst{1}
          \and
          M. Sasaki\inst{1}
          }

   \institute{Institut f\"ur Astronomie und Astrophysik, Kepler Center for Astro and Particle Physics, Eberhard Karls Universit\"at, 
              Sand 1, 72076 T\"ubingen, Germany\\
              \email{ducci@astro.uni-tuebingen.de}
              \and
              ISDC Data Center for Astrophysics, Universit\'e de Gen\`eve, 16 chemin d'\'Ecogia, 1290 Versoix, Switzerland
              \and
              INAF -- Osservatorio Astronomico di Brera, via Bianchi 46, 23807 Merate (LC), Italy
              \and 
              INAF -- Istituto di Astrofisica Spaziale e Fisica Cosmica, Via E. Bassini 15, 20133 Milano, Italy
             }

   \date{Received... ; Accepted...}

 
  \abstract
   {The transient Be/X-ray binary \src\ shows peculiar X-ray and optical variability.
    Despite numerous studies, the intrinsic properties underlying its anomalous behaviour
    remain poorly understood. 
    Since 2014 September we are conducting the first quasi-simultaneous optical and near-infrared photometric monitoring of \src\ in seven filters
    with the Rapid Eye Mount (REM) telescope, aiming to understand the properties of this binary system. 
     We found that the REM lightcurves show fast flares lasting one or two days that repeat almost 
     regularly every $\sim16.6$ days, the orbital period of the neutron star. 
     If the optical flares are powered by X-ray outbursts through photon reprocessing,
     the REM lightcurves indicate that \src\ is still active
     in X-rays: bright X-ray flares ($L_{\rm x}\gtrsim10^{37}$\,erg\,s$^{-1}$) could be observable during the periastron passages. 
     The REM lightcurves show a long-term variability that is especially pronounced in the \emph{g} band
     and decreases with increasing wavelength, until it no longer appears in the near-infrared lightcurves.
     In addition, \src\ is fainter with respect to previous optical observations
     most likely due to the higher absorption of the stellar radiation of a denser circumstellar disc.
     On the basis of the current models, we interpret these observational results with a circumstellar disc around
     the Be star observed nearly edge-on during a partial depletion phase.  
     The REM lightcurves also show short-term variability on timescales of $\sim 1$ day  
     possibly indicative of perturbations in the density distribution
     of the circumstellar disc caused by the tidal interaction with the neutron star.}


   \maketitle
%

\section{Introduction}
\label{introduction}

1A~0538$-$66 (hereafter \src) is a  Be/X-ray transient in the Large Magellanic Cloud.
It hosts a neutron star (NS) with a pulse period of 69~ms 
\citep{Skinner82} and an orbital period of $\sim 16.6$~d \citep{Johnston79}. 
\src\ was discovered in 1977 by the \emph{Ariel V} satellite \citep{White78}
during an outburst 
with peak X-ray luminosity of $8.5\times 10^{38}$~erg~s$^{-1}$,
exceeding the Eddington limit for a NS.
Other outbursts with similar luminosity were observed
with HEAO~1 \citep{Skinner80b}, and \emph{Einstein} satellites \citep{Ponman84,Skinner82}.
The X-ray outbursts of \src\ display a wide range of durations from
a few hours to $\sim 14$~days. 
The pulsations from the NS were discovered during a super-Eddington outburst
($L_x=8 \times 10^{38}$\,erg\,s$^{-1}$) observed by \emph{Einstein} satellite.
This is the only measurement of the pulse period of \src.
\emph{ROSAT}, \emph{BeppoSAX}, \emph{ASCA}, and \emph{XMM-Newton} 
observed \src\ at lower luminosities 
($L_x \approx 5 \times 10^{33} - 10^{37}$\,erg\,s$^{-1}$; 
\citealt{Campana95,Campana97,Campana02b,Corbet97,Kretschmar04}).

\src\ shows peculiar properties also in the optical band.
\citet{Skinner80} discovered bright and recurrent optical flares
(luminosity up to $\approx 3\times 10^{38}$\,erg\,s$^{-1}$, \citealt{Charles83}, hereafter \citetalias{Charles83})
with period of $\sim 16.7$\,d in phase with X-ray outbursts.
To our knowledge, this is the only high mass X-ray binary that shows fast
optical flares, associated with X-ray activity, brighter than the donor star.
Contrary to low-mass X-ray binaries, where reprocessing of X-rays into optical/UV
photons dominates or is comparable with the optical emission of the relatively faint donor star,
most of the optical emission of Be/X-ray binaries is expected to come from the bright massive donor
star and its circumstellar disc (e.g. \citealt{vanParadijs98}).
Therefore, the optical flares of \src\ represent peculiar events in the class of Be/X-ray binaries. 

The strong and complex spectroscopic and photometric variability of \src\
(see \citealt{McGowan03} for a comprehensive summary)
led to some confusion in determining the spectral type of the donor star.
It has been classified as a B7e\,II \citep{Murdin81},
OBe \citep{Pakull81}, B2\,III-IV \citepalias{Charles83}, B1\,III \citep{Hutchings85},
and B0.5\,III \citep{Negueruela02}.
The spectral type evolves to B8-9\,I during bright optical outbursts \citepalias{Charles83}.

Several models have been proposed to explain the X-ray and optical properties of \src.
In the model proposed by \citetalias{Charles83} (see also \citealt{Densham83,Smale84}),
the NS has a highly eccentric orbit ($e\geq 0.7$) around a B star.
At the periastron passage, a fraction of the material lost by the primary filling its Roche lobe
is accreted by the NS which goes into X-ray outburst.
The fraction of the material lost by the donor that is not accreted by the NS
forms a large ($R\approx 45$\,R$_\odot$ from UV spectroscopic observations, \citetalias{Charles83})
and short-lived envelope around the donor star.
In this scenario, the optical flares are powered by the reprocessing of the X-rays in
this large envelope, whose mass and size exceed
those of an ordinary Be envelope 
\citep{Maraschi83}.
The periods of inactivity (when no X-ray outbursts and optical flares are seen)
could be a consequence of the slight shrinkage of the donor star and consequent 
Roche lobe underfilling.
This is expected to be caused by temporary variations of the thermal equilibrium
of the star due to the periodic forcing of mass transfer from the outer layers
of the donor star to the NS.
\citet{Stevens88a,Stevens88b} developed an enhanced stellar wind accretion model
for \src\ where the gravitational distortion of the surface of the donor star, caused
by the periastron passage of the NS in a high eccentric orbit, leads to an enhacement of the 
mass loss rate of the donor star in the direction of the NS and consequently to much 
brighter X-ray outbursts than predicted by the standard spherically symmetric stellar winds.

The high dynamical range, spanning up to five
orders of magnitude, 
and the absence of pulsations during the outbursts with luminosity 
$L_x \leq 10^{38}$~erg~s$^{-1}$, 
have been explained by \citet{Campana95} with the source
going from direct accretion to the propeller state. 
The change of the accretion regime depends on the amount of inflowing
matter and on the magnetic field of the NS. 
\citet{Campana95,Campana02} found that the required magnetic field would be
$B\leq 2 \times 10^{11}$~G.

Another scenario was proposed by \citet{McGowan03} and \citet{Alcock01} 
(\citetalias{McGowan03} and \citetalias{Alcock01} hereafter) on the basis
of a fortuitous $\sim 5$\,yr monitoring of \src\
by the ``Massive Compact Halo Objects'' (MACHO) project
(from 1993 January 14 to 1998 May 28)
and the archival UK Schmidt and Harvard photographic B-band plates 
of taken between 1915 and 1981.
The optical lightcurve showed a high variability,
with $V$ magnitude ranging from $\sim$15 to $\sim$13
and with the brightest optical states corresponding to X-ray outbursts.
The monitoring revealed also a long-term modulation
at $P_{\rm long} = 420.8 \pm 0.8$ days (with a reddening at low fluxes)
and a short-term modulation at $P_{\rm short} = 16.6510 \pm 0.0022$ days. 
The latter confirmed the previously found orbital period. 
The orbital modulation was only seen at certain phases of the 421-d cycle, suggesting that 
the long-term modulation was related to variations in the Be star envelope.
\citetalias{McGowan03} defined two states:
\begin{enumerate}
\item \emph{quiescent:} times when the B star appears brighter ($V\sim 14.4$)
      and its optical emission is stable (i.e. no flares are seen).
      The quiescent state occurs typically in the phase range 
      $-0.06 \lesssim \phi \lesssim 0.36$ of the 421-d cycle
       (see top panel of figure 8 of \citetalias{McGowan03}).
\item \emph{active:} times when the source can be either in a low or in a high state:
     \begin{itemize}
     \item \emph{low,} times when the source is faint ($V\sim 14.8$) and no flares are seen;
     \item \emph{high,} optical/X-ray flares ($V$ magnitude up to 12.7; e.g. \citetalias{Charles83}), 
     with recurrence time of $\sim$16.6\,d. These flares are
     constrained in 421-d phase to only occur during the active state.
     \end{itemize}
\end{enumerate}

\citetalias{McGowan03} and \citetalias{Alcock01} suggested that the rise and decay of
brightness of the source (from active to quiescent state and back again)
is caused by the depletion and formation of a circumstellar disc
on the equatorial plane of the Be star, observed approximately edge-on.
The reddening at low fluxes shown by \src\ is 
in agreement with the formation of a circumstellar disc.
As the NS passes through the circumstellar 
disc, it accretes material which leads to X-ray and optical outbursts.
This scenario is consistent with the presence of Balmer emission
lines  during the active state
that indicate the presence of a circumstellar disc. 
During the quiescent state the spectrum does not show any Balmer emission,
suggesting the absence of the circumstellar disc. 
During this state, only the naked B star is presumably observed.

Recently, \citet{Schmidtke14,Schmidtke14b} studied the long-term variability of \src, from 2010 to 2014,
using the photometric I-band data of the Optical Gravitational Lensing Experiment IV (OGLE-IV) project.
They found that the source is variable in the range $I\sim15.3-15.5$.
The 421-d cycle observed by \citetalias{McGowan03} and \citetalias{Alcock01}
disappeared and long-term variations seem to occur on irregular timescales.
\citet{Schmidtke14,Schmidtke14b} searched for periodicities linking eight bright OGLE-IV flares
with four flares observed by \citet{Densham83} and obtained the ephemeris 
$P=16.6429\pm0.0007$\,d, $T_0=244\,5674.3\pm 0.2$\,JD.

Previous photometric optical monitoring of \src\ was based
on a limited number of bands.
However, a monitoring of this source in different wavelenghts would allow to gain
new information about the properties of the short-term variability
(like the fast flares that were observed at periastron)
and the long-term variability,
which would help to improve our understanding of the 
physical processes responsible for the peculiar properties of \src.
Therefore, in 2014 September we started the first daily monitoring of this source
based on quasi-simultaneous observations in seven bands with the Rapid Eye Mount (REM) telescope.
The observations and data analysis method are described in Sect. \ref{sect data analysis}.
The results are reported in Sect. \ref{sect results} and are discussed in Sect. \ref{sect discussion}.

   \begin{figure*}[ht!]
   \centering
   \includegraphics[width=12cm]{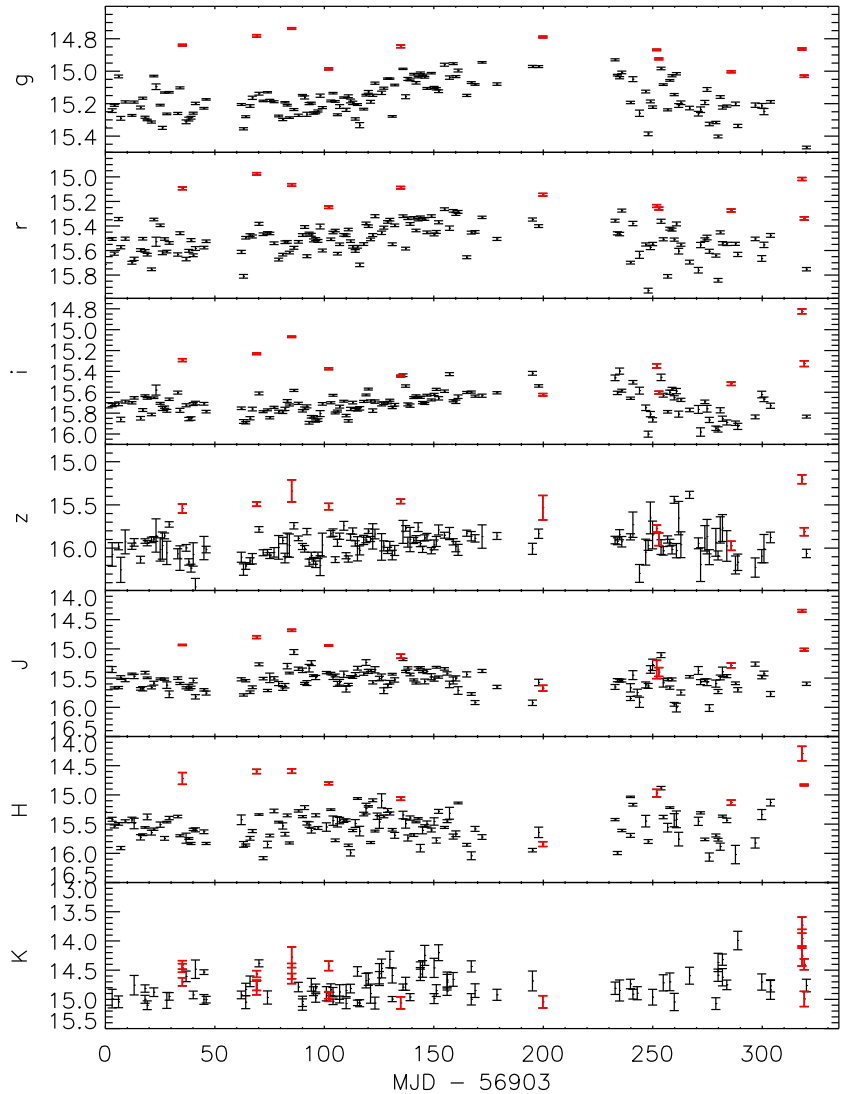}
      \caption{$g$, $r$, $i$, $z$, $J$, $H$, $K$, lightcurves of \src\ observed with REM from $T_{\rm start}=56903.2$\,MJD 
      (3 September 2014). Red points are the flares 
      lasting one or two days that repeat almost regularly every $\sim16.6$ days.
               Errors are indicated at the 68\% confidence level.}
         \label{lcrs}
   \end{figure*}

\section{Observations and data analysis}
\label{sect data analysis}

We analysed data collected with the REM telescope (\citealt{Covino04}; \citealt{Zerbi04}).
REM is a fully automatic, fast reacting telescope operated at the ESO's La Silla Observatory (Chile) since June 2004,
whose main aim is to promptly catch the afterglows of gamma-ray bursts discovered by X-ray telescopes.

REM has a Ritchey-Chretien configuration with a 60~cm primary mirror and a total $f/8$ focal ratio
and hosts the REMIR infrared camera \citep{Conconi04} and the ROSS2 optical camera \citep{Tosti04}.
The two cameras observe simultaneously almost the same field of view of $10\times10$~arcmin.
REMIR has a rotating wheel that accomodates three filters, $J$, $H$, $K$.
ROSS2 allows simultaneous imaging in the Sloan/SDSS $g$, $r$, $i$, $z$ filters
thanks to a band splitting obtained with plate dichroics.
Therefore, REM enables to observe a target simultaneously in five bands: one near-infrared band and four optical bands.

\src\ has been monitored for 257 nights (hereafter observations), 
from 2014 September 3 to 2015 July 20 (timespan $\sim 320$~d).
Each observation consists of five images (30~s integration time each) for each $g$, $r$, $i$, $z$, $J$, $H$, filter.
To avoid saturation problems, the observations in the $K$ filter consist of 15 images of 15~s integration time each.
Two sets of optical images were obtained during the $J$ and $H$ observations,
while the other three sets were captured during the $K$ observations.
For each night of observations, all the images were obtained within few minutes,
hence, for the aims of this work we can assume that the observations were quasi-simultaneous.

We performed the data analysis using the Swift Reduction Package 
(SRP)\footnote{\url{http://www.merate.mi.astro.it/utenti/covino/usermanual.html}}.
We removed images affected by high background,
bright pixels close to the target and the reference stars by visual inspection.
The images were flat-field and bias corrected, and the magnitudes of the
observed stars were obtained using the aperture photometry method.
We performed flux calibration using reference stars in the field of view
with $g$, $r$, $i$, $z$ magnitudes from \citet{Zacharias12} and $J$, $H$, $K$ magnitudes from the 2MASS catalogue \citep{Cutri03}.

For each observation and filter, the magnitude was calculated as 
the weighted mean of the magnitudes of each image.
For filter $K$ we had up to 15 images per observation. We grouped them in three sets of five images each
to obtain three weighted mean magnitudes per night.

\section{Results}
\label{sect results}

\src\ shows a high variability on short ($\sim$\,1 day)
and long (several days) timescales,
with amplitudes up to $\sim 1$ magnitude in
the optical (\emph{g}, \emph{r}, \emph{i}, \emph{z}) 
and near-infrared (\emph{J}, \emph{H}, \emph{K})
REM lightcurves (Fig. \ref{lcrs}).
Nine flares are clearly distinguishable in the $g$ band.
Since the observations in the seven bands were quasi-simultaneous, 
the corresponding measurements are indicated in red in all panels. 
These flares last one or two days and repeat almost 
regularly every $\sim16.6$ days, the orbital period of the NS, or a multiple of it.
A few exceptions are the $K$ band, where only the last flare is easily distinguishable by eye,
and the 6th and 7th flares, which are clearly observed only in the \emph{g} and \emph{r} bands.

\begin{figure}
   \centering
   \includegraphics[width=\columnwidth]{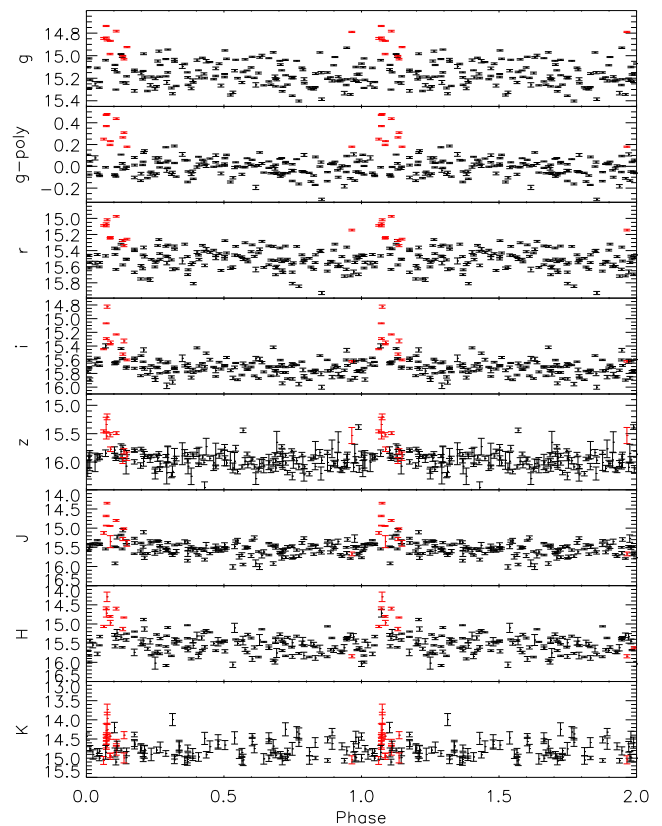}
      \caption{REM lightcurves of \src\ folded at the orbital period of \citet{Schmidtke14}.
               Errors are indicated at the 68\% confidence level. Second panel (from top) shows
               the detrended $g$ lightcurve obtained by subtracting the 5th order polynomial
               fitted to the $g$ lightcurve (excluding flares).
               Red points correspond to the flares of Fig. \ref{lcrs}.
              }
         \label{lcrs folded}
   \end{figure}

Using the ephemeris of the orbital period 
provided by \citet{Schmidtke14},
we found that the REM flares fall in the orbital phase range of $0.96-0.15$,
in agreement with the findings of \citet{Schmidtke14,Schmidtke14b}.
Figure \ref{lcrs folded} shows the REM lightcurves of \src\ folded at the orbital period value found by \citet{Schmidtke14}. 
The high scattering of the points in the first panel
is caused by the long-term variability which strongly affects this band (see below).
To emphasize the orbital modulation in the $g$ band, we plotted the detrended $g$ lightcurve obtained
from a 5th order polynomial fitted to the $g$ data without flares and then subtracted from the lightcurve
(second panel, from top, of Fig. \ref{lcrs folded}).

Flare 6 does not fall in the same phase range of the other flares ($\phi_{\rm 6th}=0.96\pm0.03$,
see Fig. \ref{lcrs folded}) and it occurs during a period in which the source was observed
sporadically (there are only two other points before the flare, then the monitoring 
was interrupted for $\sim 30$ days, see Fig. \ref{lcrs}).
Since it might not be a real flare, hereafter it will be considered separately.

Flares are associated with the periastron passage of the NS (Sect. \ref{introduction}),
therefore they should occur every $\approx 16.6$ days.
We found that in five occasions \src\ was not observed during the interval in which
the flare was expected.
In other occasions the source was covered by the monitoring
but it showed a low luminosity without any evidence of flaring activity (see Fig. \ref{lcrs folded}).
The occurrence rate of flares during the REM monitoring is thus $\approx 50$\%.

Black points of $g$, $r$, $i$, lightcurves (Fig. \ref{lcrs}) show a possible long-term variability
characterised by an increase of the flux from $t\sim 120$\,d to $t\sim 150$\,d
and a decrease from $t\sim 240$\,d to $t\sim 280$\,d. 
We used the Spearman Rank correlation test for the presence of such trend 
in the lightcurves (excluding flares).
For the $g$-lightcurve, we find a probability of no-correlation of $P_{\rm g} \simeq 0.2$\%,
while for the other bands there is no evidence for the long-term variability
($P_{\rm r}\simeq 11$\%, $P_{\rm i}\simeq 9$\%, probabilities in the range 9\% to 95\% for the other bands).
A similar trend has been observed by \citet{Schmidtke14,Schmidtke14b} with OGLE.

\section{Discussion}
\label{sect discussion}

In the previous section we presented new results about the long-term
and short-term optical variability of \src\ obtained through the REM monitoring.
In Sects. \ref{sect Long-term variability}, \ref{sect. flares},
and \ref{sect. short-term variability}, we interpret these results
on the basis of the models proposed so far for \src\
and for the dynamical evolution of circumstellar discs in Be stars.

\subsection{Long-term variability}
\label{sect Long-term variability}

The temporal evolution of the \emph{g}, \emph{r}, \emph{i} 
magnitudes in Fig. \ref{lcrs} is characterised by a low-luminosity state
when $t\lesssim 120$~d and $t\gtrsim 240$~d
and a high-luminosity state 
when $150\lesssim t \lesssim 240$~d.
The average \emph{g}, \emph{r}, \emph{i}, magnitudes computed excluding the flares 
(i.e. obtained from black points of Fig. \ref{lcrs})
during the low luminosity state observed by REM are 
$g_{\rm ll}=15.21 \pm 0.08$,
$r_{\rm ll}=15.55 \pm 0.11$,
$i_{\rm ll}=15.75 \pm 0.10$,
while during the high luminosity state they are
$g_{\rm hl}=15.04 \pm 0.07$,
$r_{\rm hl}=15.41 \pm 0.12$,
$i_{\rm hl}=15.58 \pm 0.09$.
We converted these magnitudes to Johnson \emph{V} and Kron-Cousins \emph{R} magnitudes 
using the transformation equations of \citet{Jester05}.
We obtained 
$V_{\rm ll}=15.40 \pm 0.16$, $R_{\rm ll}=15.38 \pm 0.20$ and
$V_{\rm hl}=15.24 \pm 0.16$, $R_{\rm hl}=15.20 \pm 0.20$.
The magnitudes reported by \citetalias{Alcock01} 
during the active-low state are 
$V_{\rm al}\approx 14.7-14.9$, $R_{\rm al}\approx 14.8-15$,
while during the quiescent state they are
$V_{\rm q}\approx 14.4$, $R_{\rm q}\approx 14.6-14.7$.
If we consider a systematic error of $\pm 0.1$ magnitudes
due to the zero-point uncertainty in MACHO photometry \citep{Alcock99}
a comparison between REM and MACHO magnitudes reported above suggests that 
during the REM monitoring, \src\ was fainter than in the past.
In particular, the magnitudes during the low luminosity state observed
by REM are larger than the MACHO magnitudes measured during the active-low luminosity state,
while the magnitudes during the REM high luminosity state are just slightly larger 
than those measured by MACHO during the active-low state\footnote{In addition 
to the zero-point uncertainty, another source of inaccuracy 
might lie in the conversion of magnitudes from the non-standard MACHO instrumental bandpasses
to the Johnson-Kron-Cousins system (see figure 1 of \citealt{Alcock99}).}.

   \begin{figure}
   \centering
   \includegraphics[width=\columnwidth]{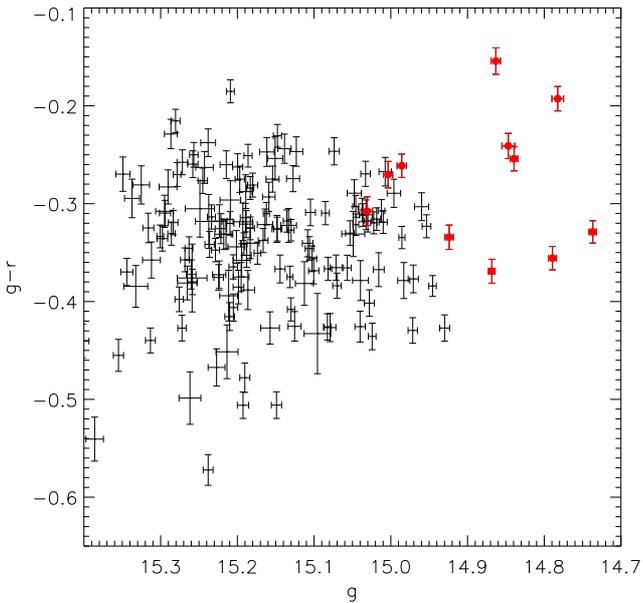}
      \caption{Colour-magnitude diagram of \src. Red points correspond to the flare of Fig. \ref{lcrs}.
              }
         \label{col-mag}
   \end{figure}

The color-magnitude diagram of \src\ (Fig. \ref{col-mag}) shows that there is not
a significant spectral variability between the flares (red points) and
the other observations (black points).
A comparison between the $g-r$ vs $g$ diagram (Fig. \ref{col-mag}) and 
the $V-R$ vs $V$ diagram of \citetalias{Alcock01} shows that the bright and bluer
branch corresponding to the quiescent state is not present in the $g-r$ vs $g$ diagram 
of the REM observations.
In the framework of the model of \citetalias{McGowan03} and \citetalias{Alcock01}, 
Fig. \ref{col-mag} suggests that a remnant of the circumstellar disc
partially covering the Be star is present during the the high-luminosity state
observed by REM, preventing to observe the naked star.

In principle, simultaneous photometric monitoring of Be stars at different wavelengths could allow 
to infer the properties of the disc and estimate the inclination angle of the system (e.g. \citealt{Carciofi06}).
In fact, fluxes at different wavelengths carry information from different regions of the circumstellar disc.
For instance, at short wavelengths (optical), most of the continuum radiation
is formed in a relatively small region near the star (e.g. within about two stellar radii for $V$ band),
whereas the emission at longer wavelengths (infrared) arises from a larger area of the disc
(e.g. roughly up to $\approx 10$ stellar radii for $K$ band).
Therefore, temporal evolution of the disc density, including the build-up and depletion
phases, as well as the mass injection properties (periodic or episodic mass injection) affect
the photometric lightcurves of different filters differently 
(\citealt{Poeckert78}; \citealt{Haubois12} and references therein).
In addition, the inclination angle of the circumstellar disc substantially modifies
the observed fluxes at different wavelengths.
\citet{Haubois12} studied the dynamical evolution of circumstellar discs around Be stars
through simulations based on a three-dimensional non-local thermal equilibrium radiative
transfer code.
They calculated the temporal evolution of photometric observables for different configurations
of the system, such as mass loss history, $\alpha$ viscosity parameter,
and inclination angle of the disc.
For several configurations of the system, they provided theoretical 
lightcurves at different wavelengths. We compared them with
the REM lightcurves of \src\ and found
a good agreement if we assume that the circumstellar disc of \src\
is observed edge-on. For this configuration of the system,
the theoretical lightcurves
at short wavelengths (figure 13 of \citet{Haubois12}; to be compared with $g$ lightcurve
of our observations) show that for inclination angles greater than $70^\circ$\footnote{\citet{Haubois12} reported
theoretical lightcurves only for inclination angles $i=0^\circ$ (face-on), $i=70^\circ$, and $i=90^\circ$ (edge-on).}
the flux decreases during the build-up phase,
reaches a plateau, and then increases during the depletion phase to the original value.
In fact, the emission in the optical bands arises from the densest inner part of the disc
and depends on the gas emission and the absorption of the stellar radiation 
in the circumstellar disc.
Therefore, if the disc is observed edge-on, during the build-up phase
the decrease of the flux caused by the increase of the absorption dominates over the 
brightening caused by the increase of the size and density of the disc. 
Similarly, during the depletion phase, the decrease of the absorption of the stellar radiation
causes an increase of the flux that reaches its maximum when the disc disappears.
If the inclination angle were $i<70^\circ$, the contribution to the optical emission 
from the disc would be dominant with respect to the absorption of the stellar radiation.
Therefore, the flux would increase during the build-up phase and decrease during the depletion phase.
The simulated lightcurves of \citet{Haubois12} predict a balance between the 
emission gas and absorption of the stellar radiation when $i=70^\circ$,
hence a  constant flux at short wavelengths is expected in this case.

The infrared (IR) bands give information about the external part of the disc
where the build-up and depletion of the disc produces the highest density variations.
Therefore, if a forming/dissipating disc is observed face-on, the IR lightcurves
are subject to high variability.
Figure 16 of \citet{Haubois12} shows that if the disc is observed edge-on,
emission and absorption balance each other and the observed flux does not vary significantly.
The absence of a significant long-term variability in the $J$, $H$, $K$, lightcurves of the REM monitoring
is thus in agreement with the edge-on scenario, and consistent with the type of variability 
observed in the $g$ band.
Based on the considerations presented at the beginning of this section,
the circumstellar disc should be always present during the REM monitoring.
Therefore, the observed long-term variability can be explained with a temporary and
partial depletion of the disc, probably caused by the tidal interactions
between the disc and the NS, which can lead in some cases to the truncation of the disc
(e.g. \citealt{Okazaki01} and \citealt{Panoglou16} for recent results about the disc truncation mechanism).
Therefore, the maximum magnitude difference between the fully developed disc state and the naked star state
must be calculated using the REM observations
($V\approx 15.4$, converted from the $g$ and $r$ bands using the transformation equations of \citealt{Jester05})
and the photospheric flux measured by \citetalias{Alcock01} ($V\approx 14.4$) during the quiescent state.
We found a value $\Delta V\approx 1$ which is 
much larger than the maximum predicted by \citet{Haubois12} for $i=90^\circ$, $\Delta V\approx 0.2$.
However, the theoretical computations of \citet{Haubois12} have been performed for isolated
Be stars.
Tidal interaction between the circumstellar disc and a NS companion
are expected to affect the density structure
of the disc (\citealt{Panoglou16} and references therein), hence deviations from the 
results of \citet{Haubois12} are expected.

\subsection{Flares}
\label{sect. flares}

In 1981 April 29, \src\ showed one of its brightest optical flares (V$\sim 12.7$).
\citetalias{Charles83} used spectro-photometric observations of the flare, 
obtained with the International Ultraviolet Explorer (IUE),
to estimate an effective temperature of $T_{\rm eff}=12500$\,K,
a radius of the emitting region of $R=45$\,R$_\odot$, 
and an optical luminosity of $L_{\rm opt}=3\times10^{38}$\,erg\,s$^{-1}$, a factor
$\sim 6.4$ brighter than in the active-low state 
($L_{\rm opt}=4.7\times10^{37}$\,erg\,s$^{-1}$, \citetalias{Charles83}).
To compare the optical luminosity of the flare reported by \citetalias{Charles83}
with those of the flares observed by REM, we fitted the spectral energy distribution (SED)
of the latter with an absorbed blackbody model. 
For the extinctions, we used the approximations of
\citet{Cardelli89} assuming for the interstellar reddening in the direction of \src\
the value $E(B-V)\approx 0.14$\footnote{The assumed value for the
interstellar reddening was extracted from the maps of \citet{Schlafly11} and \citet{Schlegel98}
available at: \url{http://irsa.ipac.caltech.edu}.}.
A significant fraction of the radiation
is expected to be emitted at UV wavelengths. Therefore, the SED obtained from the REM observations,
that covers the optical/near-IR domain, can be used to obtain only a rough estimate
of the optical luminosity during the flares.
We found that the optical luminosity during the flares is 
$2\times10^{37}\lesssim L_{\rm opt}\lesssim 15\times10^{37}$\,erg\,s$^{-1}$,
on average 40\% brighter than in the observations that preceded and followed the flares.
The excess luminosity during the flares above the underlying emission from the star and circumstellar disc
is thus $6\times10^{36}\lesssim L_{\rm opt}^{\rm exc} \lesssim 4\times10^{37}$\,erg\,s$^{-1}$.
Therefore, the flares observed by REM are much fainter than the event of 1981.
We noted that flare 6 is much brighter compared to the others: 
$L_6^{\rm exc}\approx1.7\times 10^{38}$\,erg\,s$^{-1}$.
For the reasons mentioned in Sect. \ref{sect results}, 
the results about this flare should be taken with caution.
In addition, lightcurves of Fig. \ref{lcrs} show
that flare 7 has a different spectral shape compared to the others, 
being only distinguishable from the continuum emission as flare in the optical bands.

As mentioned in Sect. \ref{introduction}, \src\ has an 
anomalously large ratio $L_{\rm opt}^{\rm exc}/L_{\rm x}$ for an early-type binary system.
Assuming the value $L_{\rm opt}^{\rm exc}/L_{\rm x}\lesssim 0.3$ reported by \citet{Maraschi83},
we found that the optical outbursts observed with REM were powered by reprocessed
X-ray flares with luminosity $L_{\rm x}\gtrsim 2-13 \times 10^{37}$\,erg\,s$^{-1}$.
\src\ was not observed in X-ray during the optical flares detected by REM.
Therefore, we cannot confirm the presence of X-ray flares simultaneous with the optical ones.

\subsection{Short-term variability}
\label{sect. short-term variability}

In addition to the periodic flares associated with the orbital period of the NS,
the REM lightcurves showed another type of short-term variability, on timescales of $\sim1$~day.
This irregular variability can be attributed to
the perturbation in the density distribution of the circumstellar disc
caused by the tidal interaction with the NS. 
Alternatively, this variability might be caused by non-radial pulsations
which can produce 
variations up to 0.1 magnitude in visual bands 
(e.g. \citealt{Percy97,Kiziloglu07} and references therein).
Since this type of variability is not irregular,
we searched for periodic signals on short timescales using the fast Lomb-Scargle
periodogram technique for unevenly sampled time series \citep{Lomb76,Scargle82}.
To avoid aliasing, we searched for signals with periods longer than the Nyquist limit
of 2~d (i.e. half of the sampling frequency of 1~d$^{-1}$).
No statistically significant periodicities associated with the observed short-term variability were
detected, thus disfavoring the ``non-radial pulsations'' hypothesis.

\section{Conclusions}
\label{sect. conclusions}

We presented the results obtained from the first daily photometric monitoring of \src\
performed quasi-simultaneously on seven bands, from optical to near-IR.
The long-term variability observed by \citet{Schmidtke14} in OGLE lightcurves 
is clearly observed also in the REM lightcurve in the $g$ band while it is absent in the near-IR bands (Fig. \ref{lcrs}).
In REM observations, \src\ is fainter than in the previous MACHO observations \citepalias{McGowan03,Alcock01},
probably because of the higher absorption of the stellar radiation by a denser circumstellar disc.
We showed that the long-term variability observed in the REM lightcurves at different wavelengths
is qualitatively in agreement with the theoretical lightcurves obtained by \citet{Haubois12}
for a circumstellar disc observed edge-on during a partial depletion phase of the disc.
Therefore, the results obtained with the REM monitoring are consistent with the scenario proposed
by \citetalias{McGowan03}, and suggest that the inclination angle of the disc should be close to $90^\circ$,
in agreement with the lower limit of $\sim 75^\circ$ obtained by \citetalias{McGowan03}.

REM observations allowed us to observe two types of short-term variability:
fast flares (with durations of 1$-$2 days)
that repeat almost regularly every $\sim16.6$ days (the orbital period of the NS),
and an irregular variability which we interpret as
due to perturbations in the density distribution
of the circumstellar disc caused by the tidal interaction with the NS.
If the optical flares are powered by X-ray outbursts through photon reprocessing (\citetalias{Charles83}; \citealt{Maraschi83}),
the observed fast flares suggest that \src\ is active in X-ray.
We made a rough estimate of the expected X-ray luminosity associated with the observed optical flares
and we found that \src\ may have reached X-ray luminosities $L_{\rm x}\gtrsim10^{37}$\,erg\,s$^{-1}$
during the REM monitoring.
We noticed that the flare 7 shows different spectral shapes compared to the other flares.
Further simultaneous observations at different wavelengths,
including X-rays and UV band, will be valuable to better characterise 
the complex observational properties shown by this peculiar binary system.

\begin{acknowledgements}
We thank the anonymous referee for constructive comments that
helped to improve the paper.
LD acknowledges Roberto Decarli for suggesting the use of REM,
Dino Fugazza for the helpful advice during the preparation of the observational strategy,
and Emilio Molinari for granting us the first six months observing time through DDT.
This work is supported by the Bundesministerium f\"ur
Wirtschaft und Technologie through the Deutsches Zentrum f\"ur Luft
und Raumfahrt (grant FKZ 50 OG 1602).
M.S. acknowledges support by the Deutsche Forschungsgemeinschaft through the Heisenberg
Fellowship SA 2131/3-1.
VD acknowledges support from Deutsches Zentrum f\"ur Luft- und Raumfahrt (DLR) through DLR-PT grant 50 OR 0702.
SM acknowledges financial contribution from PRIN INAF 2014 and from ASI/INAF Agreement  I/037/12/0
\end{acknowledgements}

\bibliographystyle{aa} 
\bibliography{538}

\end{document}